\documentclass[slac_one]{revtex4}
\usepackage{graphicx}
\usepackage{fancyhdr}
\usepackage{dcolumn}
\newcolumntype{e}[1]{D{.}{.}{#1}}

\begin{document}

\title{{\small{2005 ALCPG \& ILC Workshops - Snowmass,
U.S.A.\hfill{\normalfont SLAC-PUB-11567}}}\\ 
\vspace{12pt}
SUSY Parameter Measurements with Fittino} 

%

\author{P.~Bechtle}
\affiliation{SLAC, Stanford, CA 94025, USA}
\author{K.~Desch, P.~Wienemann} 
\affiliation{Universit\"at Freiburg,
  Physikalisches Institut, Hermann-Herder-Str.~3, D-79104 Freiburg,
  Germany}

    
\begin{abstract}
  This article presents the results of a realistic global fit of the
  Lagrangian parameters of the Minimal Supersymmetric Standard Model
  with no assumptions on the SUSY breaking mechanism using the fit
  program Fittino. The fit is performed using the precision of future
  mass measurements of superpartners at the LHC and mass and polarized
  topological cross-section measurements at the ILC. Higher order
  radiative corrections are accounted for wherever possible to date.
  Results are obtained for a modified SPS1a MSSM benchmark scenario
  (general MSSM without assumptions on the breaking mechanism) and for
  a specific mSUGRA scenario. Exploiting a simulated annealing
  algorithm, a stable result is obtained without any {\it a priori}
  assumptions on the fit parameters. Most of the Lagrangian parameters
  can be extracted at the percent level or better if theoretical
  uncertainties are neglected. Neither LHC nor ILC measurements alone
  will be sufficient to obtain a stable result.
\end{abstract}

\maketitle

\thispagestyle{fancy}



\section{Introduction}\label{sec:intro}

Provided low-energy Supersymmetry (SUSY)~\cite{susy} is realized in
Nature, the next generation of colliders, the Large Hadron Collider
(LHC)~\cite{lhc} and the International Linear Collider
(ILC)~\cite{ilc} are likely to produce most particles of the SUSY
spectrum and will allow for precise measurements of their properties.
If SUSY is established experimentally, it is the main task to explore
the unknown mechanism of SUSY breaking.  Specific SUSY breaking models
(e.~g.~minimal supergravity (mSUGRA)~\cite{msugra}) can be tested
against the observables in a straight-forward manner due to the small
number of parameters. However, an exploration of the parameters of the
general Minimal Supersymmetric Standard Model (MSSM) parameter space
is significantly more ambitious, especially without the use of any
{\it a priori} assumptions on the parameter values. Using the
best-studied MSSM scenario available to date,
SPS1a~\cite{ref:SPS,ref:SPA,ref:LHCILC}, the prospects for SUSY
parameter measurements using the fit program
Fittino~\cite{ref:FittinoProgram} are outlined in the following.


\section{SUSY Parameter measurement in the MSSM with Fittino}\label{sec:sps1a}

Fittino has been created to determine the parameters of the MSSM
Lagrangian without any {\it a priori} assumptions, using available
loop-level precision predictions (e.~g.~\cite{ref:SPheno}) and
observables from present and future colliders, cosmology and rare
decay measurements in a $\chi^2$ fit. Since finding the
$\chi^2$-minimum in a many-dimensional ($\approx 20$) parameter space
is difficult, a 3-step technique is used: First, the parameters are
estimated using tree-level relations. Second, simulated annealing is
used to find the global minimum. Third, a global fit is used to find
the precise minimum and determine the parameter uncertainties and
correlations.

In order to test the parameter measurement using simulated
measurements from future colliders, the SPS1a' scenario~\cite{ref:SPA}
is chosen.  Unification is assumed in the first two generations, and
the top quark mass $m_{t}$ is fitted as an additional parameter to
account for parametric uncertainties. In total, 19 free parameters are
fitted. The observables and their simulated uncertainties used in the
fit are given in~\cite{ref:LHCILC,ref:FittinoPhysics}. By fitting to
edges in LHC spectra instead of masses derived from the
spectra~\cite{ref:LHCILC}, correlations among observables are taken
into account where known to date.

\begin{figure*}[t]
  \centering
  \includegraphics[width=0.40\textwidth]{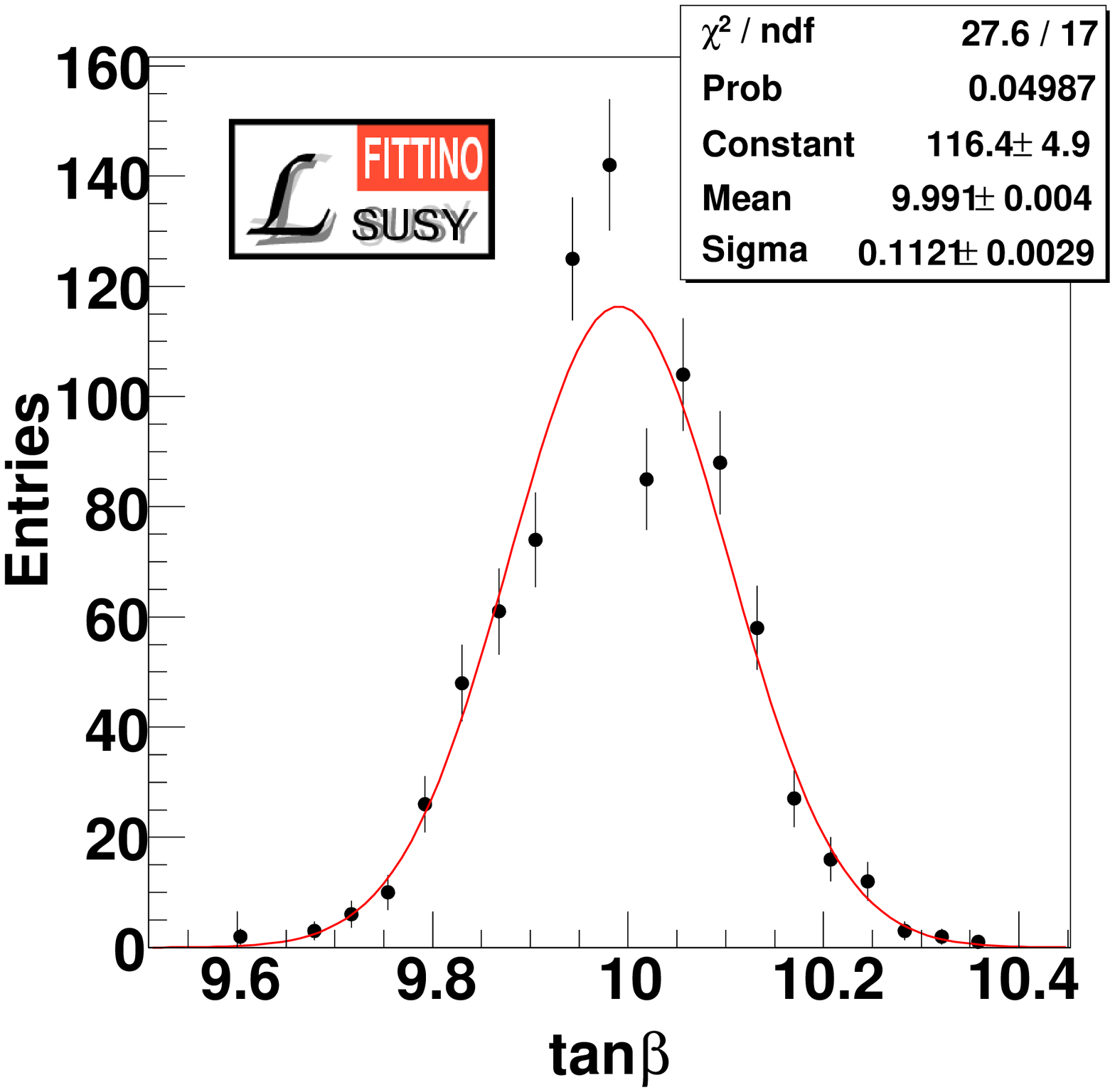}
  \includegraphics[width=0.417\textwidth]{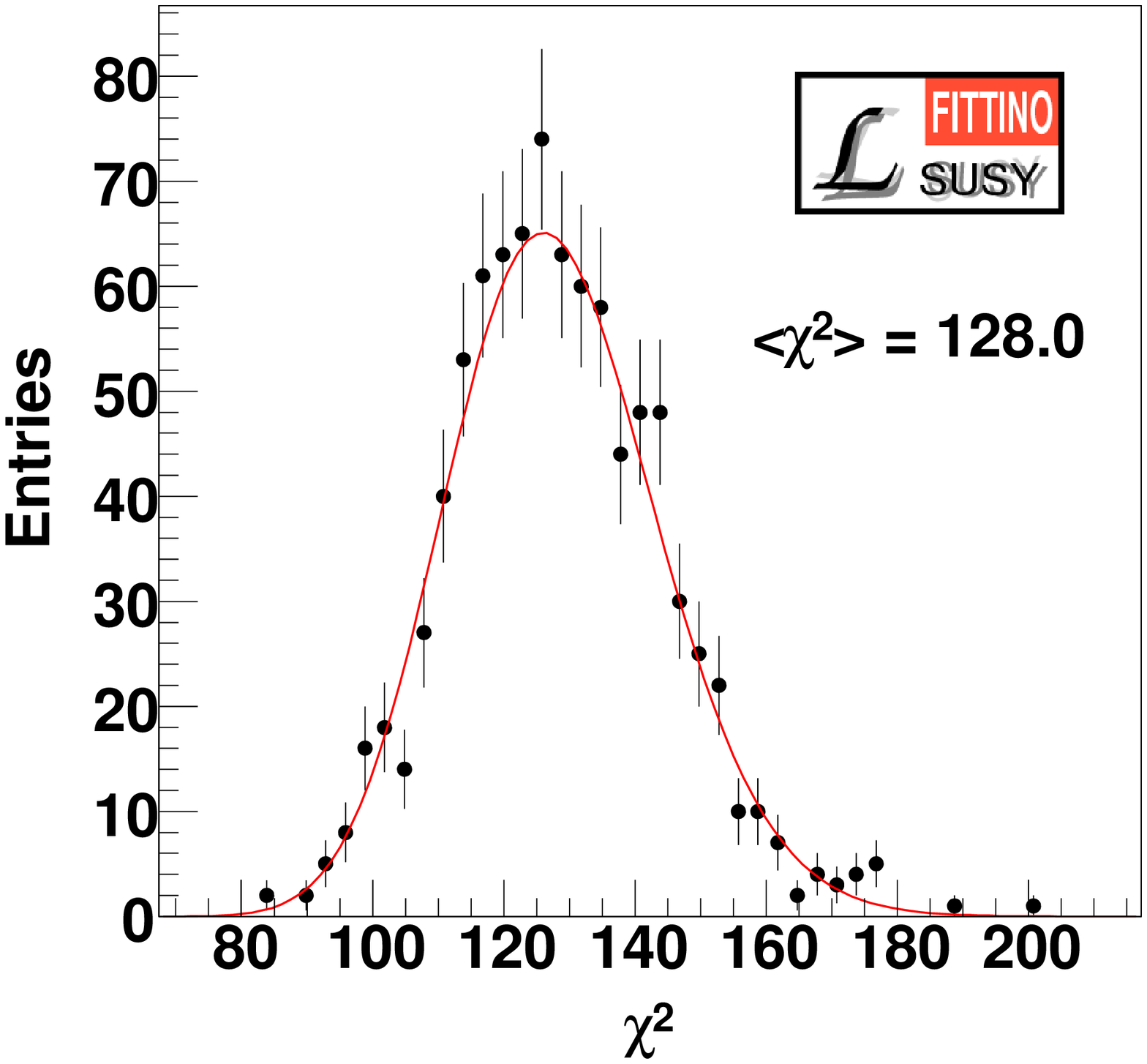}\vspace{-7mm}
  \caption{{\sl
      The plot on the left shows the toy fit value distributions for
      $\sim$1000 independent fits with observables smeared within their
      uncertainties for the parameter $\tan\beta$. The right plot shows the
      $\chi^2$ distribution for $\sim$1000 independent fits with
      observables smeared within their uncertainties. The mean $\chi^2$
      from a $\chi^2$ function fitted to the observed distribution of
      128.0 agrees with the expectation of 129.0 $\pm$ 0.7~\cite{ref:FittinoPhysics}.}}
  \label{fig:pulldist}
\end{figure*}

Using simulated measurements from LHC only, most relative
uncertainties on the parameters are in the order of 100\,\%.
Therefore, the data from the ILC is needed to understand the
low-energy MSSM Lagrangian. The fit including the simulated ILC
measurements is stable, very well under control (see
Fig.~\ref{fig:pulldist}) and shows no biases~\cite{ref:FittinoPhysics}. 

\begin{table}[t]
  \begin{center}
    \begin{tabular}{l e{7} e{7} e{7} | l e{7} e{7} e{7}}
      \hline
      \multicolumn{1}{c}{Parameter} & 
      \multicolumn{1}{c}{Fit value} & \multicolumn{1}{r}{Uncertainty} &
      \multicolumn{1}{c|}{Uncertainty} &
      \multicolumn{1}{c}{Parameter} & 
      \multicolumn{1}{c}{Fit value} & \multicolumn{1}{r}{Uncertainty} &
      \multicolumn{1}{c}{Uncertainty} 
      \\
      & \multicolumn{1}{c}{\& ``True'' Value} & \multicolumn{1}{c}{(exp.)} & \multicolumn{1}{c|}{(exp.+theor.)} &
      & \multicolumn{1}{c}{\& ``True'' Value} & \multicolumn{1}{c}{(exp.)} & \multicolumn{1}{c}{(exp.+theor.)}
      \\
      \hline
      $\tan\beta$          &  10.00                     & 0.11 & 0.15                    & 
      $M_{\tilde{q}_R}$    &  503.   \;\text{GeV}       & 12.  \;\text{GeV} &  16.  \;\text{GeV}   \\
      $\mu$                &  400.4  \;\text{GeV}       & 1.2\;\text{GeV} & 1.3\;\text{GeV}          & 
      $M_{\tilde{b}_R}$    &  497.   \;\text{GeV}       & 8.  \;\text{GeV} &  16.  \;\text{GeV}    \\
      $X_{\tau}$           &  -4449.  \;\text{GeV}      & 20.\;\text{GeV} & 29. \;\text{GeV}         & 
      $M_{\tilde{t}_R}$    &  380.9   \;\text{GeV}      & 2.5  \;\text{GeV} &  3.7    \;\text{GeV} \\
      $M_{\tilde{e}_R}$    &  115.60  \;\text{GeV}      &  0.13 \;\text{GeV} & 0.43   \;\text{GeV}   & 
      $M_{\tilde{q}_L}$    &  523.   \;\text{GeV}       &  3.2 \;\text{GeV} & 4.3    \;\text{GeV}  \\
      $M_{\tilde{\tau}_R}$ &  109.89  \;\text{GeV}      &  0.32  \;\text{GeV} & 0.56    \;\text{GeV} & 
      $M_{\tilde{t}_L}$    &  467.7   \;\text{GeV}      & 3.1 \;\text{GeV} & 5.1    \;\text{GeV}   \\
      $M_{\tilde{e}_L}$    &  181.30  \;\text{GeV}      &  0.06  \;\text{GeV} & 0.09    \;\text{GeV} & 
      $M_1$                &  103.27 \;\text{GeV}       & 0.06  \;\text{GeV} & 0.14   \;\text{GeV} \\
      $M_{\tilde{\tau}_L}$ &  179.54  \;\text{GeV}      &  0.12  \;\text{GeV} & 0.17    \;\text{GeV} & 
      $M_2$                &  193.45 \;\text{GeV}       & 0.08  \;\text{GeV} & 0.13  \;\text{GeV}  \\
      $X_{\text{t}}$     &  -565.7  \;\text{GeV}        & 6.3   \;\text{GeV} & 15.8   \;\text{GeV}   & 
      $M_3$                &  569.  \;\text{GeV}        &  7.  \;\text{GeV} &   7.4   \;\text{GeV} \\
      $X_{\text{b}}$  &  -4935. \;\text{GeV}            & 1207. \;\text{GeV} & 1713.  \;\text{GeV}   & 
      $m_{\text{A}_{\text{run}}}$ & 312.0 \;\text{GeV}  & 4.3  \;\text{GeV} & 6.5   \;\text{GeV}   \\
       &                     &                   &                                                   &  
       $m_{\text{t}}$     &  178.00  \;\text{GeV}        &  0.05  \;\text{GeV} & 0.12  \;\text{GeV} \\
      \hline
    \end{tabular}\vspace{-3mm}
  \end{center}
  \caption{\sl The Fittino fit result for the SPS1a' inspired scenario.
    The left column shows the values predicted by SPheno version 2.2.2
    for this scenario, which for all parameters are identical to the central values of
    the fit with unsmeared input observables, the third column displays
    the parameter uncertainties for the fit with experimental
    uncertainties only.  The last column shows the parameter
    uncertainties for the fit with experimental and theoretical
    uncertainties~\cite{ref:FittinoPhysics}.}
  \label{tab:UnsmearedFitResultSPS1aPrime}
\end{table}

The numerical values of the fit uncertainties are given in
Tab.~\ref{tab:UnsmearedFitResultSPS1aPrime}. Both with and without the
inclusion of present theoretical uncertainties, the precision for most
of the parameters lies in the per-cent or sub-percent range.
Theoretical uncertainties, however, can increase the uncertainties by
up to a factor of 2.5 (in the case of $X_t$) and hence need to be
reduced to make use of the full experimental information of the
high-precision ILC measurements.


\section{SUSY Parameter Measurement in mSUGRA}\label{sec:msugra}

\begin{table}
  \begin{center}
    \begin{tabular}{l e{7} e{7} e{7} e{7} e{7}}
      \hline
      & \multicolumn{1}{l}{SPS1a' value} & \multicolumn{1}{l}{Fitted value}
      & \multicolumn{1}{l}{$\Delta_{\text{LHC$+$ILC}}$} & 
      \multicolumn{2}{l}{$\Delta_{\text{LHC only}}$}\\ \hline
      $\tan \beta$   &   10.000 &   10.000 & 0.036 &  1.3\\
      $M_0$ (GeV)    &   70.000 &   70.000 & 0.070 &  1.4\\
      $M_{1/2}$ (GeV)&  250.000 &  250.000 & 0.065 &  1.0\\
      $A_0$ (GeV)    & -300.0   & -300.0   & 2.5   & 16.6\\
      \hline
    \end{tabular}\vspace{-3mm}
  \end{center}
  \caption{\sl Fit results for fits within mSUGRA scenario. The meanings of
    the columns are (starting from the left): SPS1a' values, the fitted mSUGRA
    parameters, parameter uncertainties for a fit to LHC$+$ILC observables 
    and parameter uncertainties for a fit to ``LHC only'' 
    observables. In both cases theoretical uncertainties are not included~\cite{ref:FittinoPhysics}.}
  \label{tab:mSUGRAFitResults}
\end{table}

High-scale MSSM scenarios such as mSUGRA with only 4 real parameters can
also be tested using the limited set of measurements from the LHC
alone. A result of such a fit is shown in
Tab.~\ref{tab:mSUGRAFitResults}~\cite{ref:FittinoPhysics}. However,
also in such a case the precision obtained with the combined LHC and
ILC data is more precise by more than an order of magnitude, and hence
the sensitivity on deviations from a pure high-scale scenario is
strongly enhanced by the ILC, underlining its ability to be a
telescope to GUT scale physics.

\section{Conclusions}\label{sec:summary}

Using Fittino, the Lagrangian parameters of the MSSM, assuming real
parameters but without assumptions on the SUSY breaking mechanism, can
be correctly reconstructed without usage of {\it a priori}
information.  This has been achieved using simulated precision
measurements at the LHC and ILC as input to a global fit exploiting
the techniques implemented in the program Fittino.  Most of the
Lagrangian parameters can be determined to a precision around the
percent level. For some parameters an accuracy of better than 1~per
mil is achievable if theoretical uncertainties are neglected. Assuming
present theoretical uncertainties for the predictions, the precision
of the Lagrangian parameter measurement is significantly deteriorated.
While the experimental information from the LHC alone without the
additional information from the ILC is not sufficient to determine the
full low-energy MSSM Lagrangian parameters, it is enough to determine
the parameters of a high-scale mSUGRA scenario. However, the precision
of the parameter determination (and hence the sensitivity on
deviations from perfect unification) increases by more than an order
of magnitude if ILC data are added. These results highlight the need
of the ILC for a precise understanding of Supersymmetry and its value
to explore physics from the Terascale to the GUT scale..

%
%

\end{document}